\documentclass[11pt,a4paper]{article}
\usepackage{graphicx}
\usepackage{amsmath,amssymb}
\usepackage{natbib}
\usepackage{float}
\usepackage{makecell}
\usepackage[table,dvipsnames]{xcolor}

\usepackage[T1]{fontenc}
\usepackage{lmodern}
\usepackage{microtype}
\usepackage[margin=1in]{geometry}
\definecolor{Primary}{HTML}{156082} 
\definecolor{Accent}{HTML}{E97132}  
\definecolor{Link}{HTML}{467886}    
\colorlet{Soft}{Primary!10}         

\usepackage[colorlinks=true,linkcolor=Accent,citecolor=Primary,urlcolor=Link]{hyperref}

\usepackage{titlesec}
\titleformat{\section}{\Large\bfseries\color{Primary}}{\thesection}{0.8em}{}
\titleformat{\subsection}{\large\bfseries\color{Accent}}{\thesubsection}{0.6em}{}
\titleformat{\subsubsection}{\normalsize\bfseries}{\thesubsubsection}{0.6em}{}

\usepackage[most]{tcolorbox}
\tcbset{enhanced,sharp corners,boxrule=0pt,frame hidden}
\newtcolorbox{highlightbox}[1][]{colback=Soft,borderline west={2pt}{0pt}{Primary},left=10pt,right=10pt,top=8pt,bottom=8pt,#1}
\newtcolorbox{accentbox}[1][]{colback=Soft,borderline west={2pt}{0pt}{Accent},left=10pt,right=10pt,top=8pt,bottom=8pt,#1}

\usepackage{soul} 
\sethlcolor{Soft}

\newcommand{\dfn}[1]{\textbf{\textcolor{Accent}{#1}}}

\usepackage{booktabs}
\usepackage{enumitem}
\setlist{itemsep=0.25em,topsep=0.25em}

\usepackage{titling}

\pretitle{\begin{center}\LARGE\bfseries\color{black}\vspace{-1em}}
\posttitle{\par\end{center}\vspace{-1em}}  

\preauthor{\begin{center}\normalsize\vspace{-0.5em}}
\postauthor{\par\end{center}\vspace{-1em}} 

\date{}

\title{Reframing Three-Dimensional Morphometrics Through Functional Data Innovations}

\author{Aneesha Balachandran Pillay\textsuperscript{a},
Issa-Mbenard Dabo\textsuperscript{b}
        Sophie Dabo-Niang\textsuperscript{c},
        and Dharini Pathmanathan*\textsuperscript{a,d,e}}

\date{} 

\begin{document}
\maketitle
\vspace{-4em} 

\noindent\textsuperscript{a}\,Institute of Mathematical Sciences, Faculty of Science, Universiti Malaya, 50603 Kuala Lumpur, Malaysia\\
\textsuperscript{b}\,Mathematics, Division of Science, New York University Abu Dhabi, United Arab Emirates\\
\textsuperscript{c}\,Laboratoire Paul Painlev\'e UMR CNRS 8524, INRIA--MODAL, Universit\'e de Lille, France\\
\noindent\textsuperscript{d}\,Universiti Malaya Centre for Data Analytics, Universiti Malaya, 50603 Kuala Lumpur,
Malaysia\\
\noindent\textsuperscript{e}\,Center of Research for Statistical Modelling and Methodology, Faculty of Science,
Universiti Malaya, 50603 Kuala Lumpur, Malaysia\\
\vspace{0.6em}
\noindent\textit{*Corresponding author:} dharini@um.edu.my

\begin{abstract}
This study innovates geometric morphometrics by incorporating functional data analysis, the square-root velocity function (SRVF), and arc-length parameterisation for 3D morphometric data, leading to the development of seven new pipelines in addition to the standard geometric morphometrics (GM) approach.. This enables three-dimensional images to be examined from perspectives that do not neglect curvature, through the combined use of arc-length parameterisation, soft-alignment, and elastic-alignment. A simulation study was conducted to demonstrate the general effectiveness of eight pipelines: geometric morphometrics (GM, baseline), arc-GM, functional data morphometrics (FDM), arc-FDM, soft-SRV-FDM, arc-soft-SRV-FDM, elastic-SRV-FDM, and arc-elastic-SRV-FDM. These pipelines were also applied to distinguish dietary categories of kangaroos (omnivores, mixed feeders, browsers, and grazers) using cranial landmarks obtained from 41 extant species. Principal component analysis was conducted, followed by classification analysis using linear discriminant analysis, multinomial regression and support vector machines with a linear kernel. The results highlight the effectiveness of functional data analysis, together with arc-length and SRVF-based approaches, in opening the door to more robust perspectives for analysing three-dimensional morphometrics, while establishing geometric morphometrics as the baseline for comparison.
\end{abstract}

\begin{quotation}
\noindent\textbf{Keywords:} Arc-length parameterisation; elastic alignment; functional data analysis; geometric morphometrics; square-root velocity function
\end{quotation}

\section{Introduction}
Capturing three-dimensional shape variation is crucial for disciplines spanning evolutionary biology to medical diagnostics. Over the years, a variety of shape parameterisation strategies have been developed, including landmark-based models, boundary point distributions (e.g., semi-landmarks and outlines), skeletal models, functional representations, and topological descriptors. Each has distinct advantages, such as boundary models that emphasise object contours, skeletal models that compactly encode internal structure, and topological approaches that provide scale-invariant insights. Nevertheless, landmark-based models remain dominant in geometric morphometrics due to their intuitive interpretation and compatibility with anatomical homology \cite{Drydmard16,zelditch2004geometric}.\\
In classical statistical shape analysis, landmark configurations are typically aligned using generalised Procrustes analysis (GPA) to remove translation, rotation, and scale, followed by principal component analysis (PCA) to describe variation \cite{Drydmard98,webster2010practical}. These Procrustean methods treat aligned configurations as elements of a Euclidean space, enabling standard multivariate statistical techniques. Such extrinsic approaches perform inference in a linearised setting rather than respecting the nonlinear geometry of the shape manifold \cite{patrangenaru2015nonparametric}.

Intrinsic methods recognise that shape spaces are typically non-Euclidean and conduct inference directly on the manifold using geodesic distances, Fr\'echet means, and Riemannian metrics \cite{bhattacharya2012nonparametric,patrangenaru2015nonparametric}. These are especially relevant when the curvature of the shape space affects relationships, such as in closed curves or deformable 3D objects. However, reducing continuous curves or surfaces to sparse sets of landmarks inevitably discards fine-scale detail and may introduce bias when variation is large or the manifold is strongly curved \cite{kent1997consistency,huckemann2010intrinsic}. These limitations have prompted the development of functional data alignment-invariant models and methods that better adhere to the geometry of shape space.

Functional data analysis (FDA) presents then a promising alternative by modeling curves as smooth functions within a functional space \cite{ramsay2005functional}, treating the shape as a realisation of a stochastic process rather than a set of discrete coordinates.  Srivastava and Klassen (2016)\cite{SrivastavaKlassen2016} motivate the need for rigorous statistical tools to analyse functions and shapes while handling invariances such as translation, rotation, scaling, and re-parameterisation. Their geometric framework unifies registration and shape analysis, enabling tools such as functional principal component analysis (FPCA) for vector-valued functions. Wu et al. (2024)\cite{wu2024shape} presents a framework that integrates geometric shape analysis with functional data methods to better capture structural features of curves, such as peaks and valleys, beyond what traditional based approaches offer. The framework developed tools for shape fitting, shape-fPCA, and shape regression, extending these ideas to manifold-valued functions and demonstrating their effectiveness on real and simulated datasets.

A critical consideration when modeling curve-like shapes is the choice of parameterisation. Arc-length parameterisation enables consistent assessment of complex-shaped signals, including hysteretic load-unload curves, by eliminating variability due to uneven sampling or velocity \cite{hartlen2022arclength}. In shape analysis, the space of unparameterised curves is typically modeled as a quotient of parameterised curves under the action of the reparameterisation group, and arc-length parameterisation is often used as a canonical representative to identify each equivalence class, leading to uniformly sampled and geometry-preserving shape comparisons \cite{Tumpach2017}.

Functional data morphometrics (FDM) represents landmark trajectories as multivariate functions, allowing detailed deformation patterns and feature-based registration \cite{guo2022data,ramsay1998curve}. Multivariate functional principal component analysis (MFPCA) extends landmark trajectories to multi-dimensional trajectories, especially when temporal or arc-length ordering is essential \cite{happ2018multivariate,yao2005functional}. Parallel advances in elastic shape analysis, such as the square root velocity function (SRVF) framework, leverage the Fisher--Rao Riemannian metric to separate amplitude and phase variation, aligning curves to a Karcher mean template \cite{srivastava2011registration}. This manifold-aware approach is part of a broader object-oriented data analysis perspective \cite{marron2021object} that integrates geometric structure with elastic deformation metrics. Intrinsic methods offer a theoretically robust enhancement to Procrustean techniques, particularly in the context of high-dimensional or complex shape data. Katina et al. (2016)\cite{katina2021functional} illustrated the application of functional data analysis to 3D facial surfaces by treating each coordinate as a function over a 2D grid and employing FPCA to encapsulate the primary variations in shape. Drawing inspiration from this functional paradigm, our study adopts a similar strategy for examining 3D landmark trajectories or curves, utilizing arc-length parameterisation and MFPCA to more effectively analyse anatomical shape variation. This paper introduces two traditional geometric morphometric methods alongside six functional data-based geometric morphometric pipelines for three-dimensional data:

\begin{enumerate}
 \item \dfn{GM:}  a classical geometric morphometrics (GM) approach was used to remove non-shape information by applying Geometric Procrustes  Analysis (GPA)
  \item \dfn{Arc-GM:} reparameterises each shape  to uniform arc before applying GPA
  \item \dfn{FDM}: a three-dimensional extension of Pillay et al. (2024)\cite{pillay2024functional} which models each three-dimensional outline as a smooth multivariate functional data and applies multivariate functional principal component analysis on the raw parameterisation.
  \item \dfn{Arc-FDM}: reparameterises each functional data (curve) to uniform arc length before spline-basis smoothing and analysis to ensure even sampling.
 
  \item \dfn{Soft-SRV-FDM}: blends the identity mapping with an estimated SRVF warp, achieving a soft elastic alignment that balances global outline preservation with the ability to capture fine-scale shape variation.
  
  \item \dfn{Arc-Soft-SRV-FDM}: reparameterises each shape  to uniform arc before applying Soft-FDM-SVR
  \item \dfn{Elastic-SRV-FDM}: applies full SRVF-based elastic alignment, estimating and applying the optimal warping function to remove phase variation, thereby isolating amplitude differences and capturing shape-related variation.
  \item \dfn{Arc-Elastic-SRV-FDM}: reparameterises each shape  to uniform arc before applying elastic-SVR-FDM
\end{enumerate}

\begin{figure}[H]
  \centering
  \includegraphics[width=\textwidth]{\detokenize{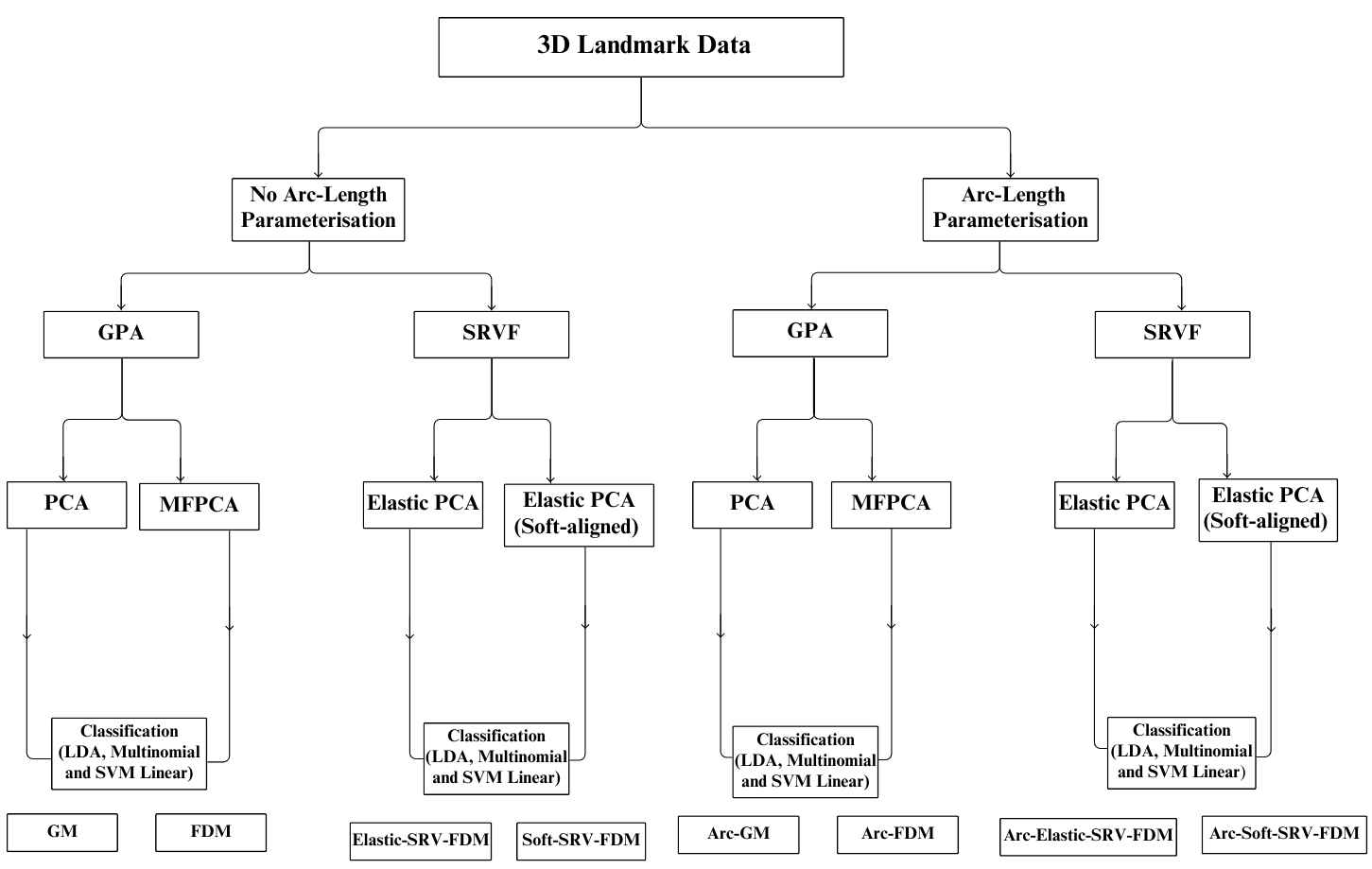}}
  \caption{Flowchart of the eight pipelines combining functional data analysis and geometric morphometrics for three-dimensional morphometrics data.}\label{fig::pipelines}
\end{figure}

Figure~\ref{fig::pipelines} illustrates the flowchart of the eight pipelines that integrate functional data analysis and geometric morphometrics for three-dimensional morphometric data.

Of these approaches, the final seven represent innovations derived from existing methods found in the literature.
Following the aforementioned geometric analyses, the results of each method are classified utilising three approaches: linear discriminant analysis (LDA), linear support vector machines (SVM), and multinomial logistic regression. These are trained on the extracted principal component scores from each method's data. The pipelines previously mentioned were evaluated using both a simulation study and a real-world dataset consisting of 3D kangaroo skull landmarks \cite{butler20213d}, illustrating significant biological diversity across various dietary categories.

Relative to traditional geometric morphometrics techniques, the functional data framework presents a cohesive and versatile paradigm that effectively accommodates both aligned and unaligned shape representations. It preserves significant phase variation through arc-parameterisation and SRVF alignment, while it also improves classification accuracy through incorporation with supervised learning models. This study elucidates the balance between alignment strategies and classification accuracy by providing both quantitative metrics and visual evaluations of reconstruction fidelity. Our findings are intended to offer practical guidance for the customisation of geometric morphometrics pipelines in accordance with specific research objectives, whether in the fields of taxonomy, evolutionary analysis, or medical morphology.

This work is organised as follows. Section~\ref{sec:methods} details the methods based on the GM and FDM frameworks. Section~\ref{sec:sim} presents a simulation study to compare FDM with its classical counterpart GM. Section~\ref{sec:app} describes the application to 3D kangaroo landmark data. Section 5 provides the discussion for both the simulation study and empirical data. Section~\ref{sec:conc} concludes. 

\section{Methodology} \label{sec:methods}


 Let 
\[
\mathbf{x}_k = (x_{k,\tau_1}, \\ \ldots, \\ x_{k,\tau_N}), 
\quad
\mathbf{y}_k = (y_{k,\tau_1}, \\ \ldots, \\ y_{k,\tau_N} ), 
\quad
\mathbf{z}_k = ( z_{k,\tau_1}, \\ \ldots, \\ z_{k,\tau_N} ),
\]
$
\quad k = 1, \dots, n,$
be the symmetric shape landmark coordinates for $n$ specimens.  
Here, $\tau_1, \dots, \tau_N$ are the observed landmark points, and $N$ is the number of landmarks on a dimensional domain (Figure~\ref{fig::reconplot}) based on the crania of kangaroos Figure S1).

\subsection{Generalised Procrustes  Analysis}

In this analytical pathway, we performed  GPA. A curve-sliding matrix was provided to guide the sliding of semi-landmarks along tangent directions, minimising bending energy based on the thin-plate spline criterion. 
After GPA, averaged configurations were computed from the aligned coordinate array and stored. Visualisations, including grand mean shapes and reference-to-target comparisons, were used to assess alignment quality. Given the bilateral symmetry of kangaroo crania, a bilateral symmetry analysis was applied. A landmark pairing file specified corresponding bilateral landmarks. Symmetrised shape configurations were generated using object symmetry (not side or replicate symmetry), and the output was used in subsequent analyses. 
From the symmetrised GPA-aligned coordinates, shape variables were extracted by computing species-level means, corresponding to unique specimen identifiers. The symmetrised shape data were aggregated and reshaped into a 3D array for downstream shape-space analysis and visualisation.
In parallel, centroid size, representing overall cranial scale, was averaged across replicates for each specimen and subsequently across species. This provided a size vector aligned with the shape configurations.



Generalised Procrustes analysis was conducted using the gpagen() function from the geomorph package in R \cite{adams2013geomorph}, which optimally aligns configurations by minimizing the summed squared distances between corresponding landmarks.
To improve alignment fidelity for curves and complex outlines, semi-landmarks were allowed to slide along tangents to the curve using thin-plate spline (TPS) bending energy minimisation, a standard method for minimising landmark placement artifacts while preserving biological signal \cite{gunz2005semilandmarks,guo2022data}. This was particularly relevant for cranial curves where homologous correspondence is geometrically approximate rather than anatomically fixed.
After GPA and sliding, the aligned landmark data were analysed using classical PCA to reduce dimensionality and identify dominant modes of variation. The principal components that accounts for $95\%$ of the variation explained were retained for further analysis. 


\subsection{Arc-Length Parameterisation}

The arc-length parameterisation reparametrises a curve so that the speed is constant (usually 1).
Given a landmark curve $f(t)$ with $t \in T=[0,1]$, its arc-length is:

$$
s(t) = \int_0^t \|\dot{f}(\tau)\| \, d\tau,
$$
where $\dot{f}(t)$ is the derivative of $f$ with respect to $t$ and $\|\cdot\|$ is the Euclidean norm in $\mathbb{R}^d$, $d\ge 1$.
The total length is
$
L = s(1).
$

Arc-length parameterisation means defining a new parameter $u \in [0, L]$ such that:

$$
g(u) = f(t(u)), \quad \text{with} \quad \left\|\frac{dg}{du}\right\| = 1.
$$

Here, $t(u)$ is the inverse of $s(t)$.

Arc-length reparameterisation presents numerous benefits: it mitigates the influence of variable speed in curve comparisons and guarantees uniform distribution along the curve.

\subsection{Functional Data Framework}

Landmark registration offers a straightforward approach to detect and align specific data points for each observation to the corresponding mean value. This provides a better representation of the mean in terms of amplitude variation. Each observation is vector-valued, as three spatial coordinates the $x$, $y$, and $z$ coordinates are involved.

\begin{figure}[H]
  \centering
  \includegraphics[width=0.8\textwidth]{\detokenize{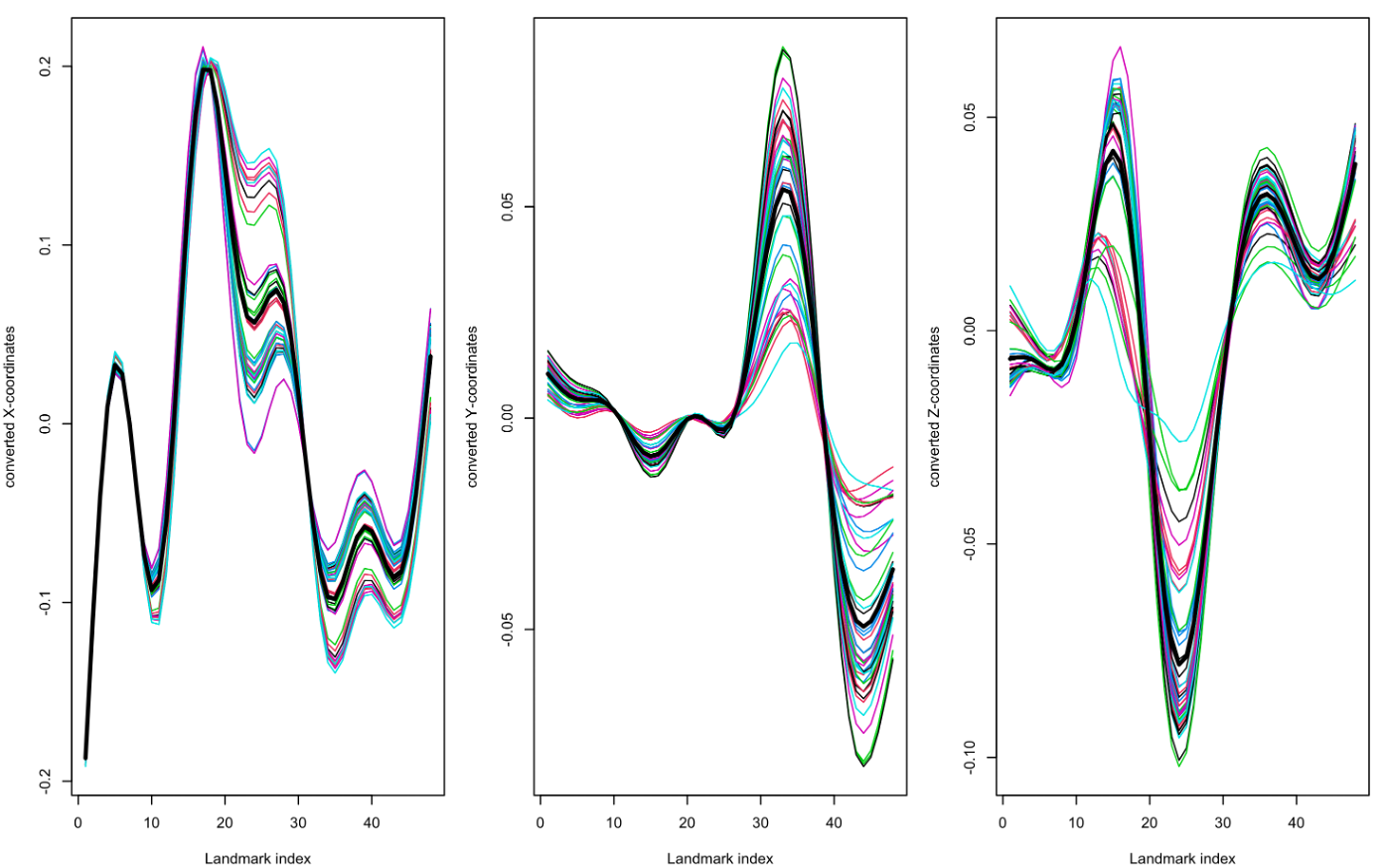}}
  \caption{Functional representations of kangaroo landmark data across three coordinate dimensions using the FDM approach. Each coloured line corresponds to one specimen.}\label{fig:kangarooxyz}
\end{figure}

To implement functional data in an object-oriented way, the raw data is converted into functions. For example, for the $x$-coordinates, using the observation points (boundaries) $\tau_1, \dots, \tau_N$ and the set of discrete raw data (landmark values), a univariate functional data sample
\[
\{ X^{(1)}(\cdot), \dots, X^{(n)}(\cdot) \},
\]
is constructed. This procedure is also applied to the $y$-coordinates and $z$-coordinates, yielding the functional samples
\[
\{ Y^{(1)}(\cdot), \dots, Y^{(n)}(\cdot) \}
\quad \text{and} \quad
\{ Z^{(1)}(\cdot), \dots, Z^{(n)}(\cdot) \},
\]
respectively (see for instance Figure~\ref{fig:kangarooxyz}).

These three univariate functional datasets form a multivariate functional dataset, with $n$ outlines, each yielding a vector of $n$ observations defined over a 3-dimensional functional domain.  
The \texttt{MFPCA} package~\citep{happ2018multivariate} is used to perform the conversion to functional data~\citep{ramsay2005functional}.

\subsubsection{Multivariate Functional Principal Component Analysis}

After acquiring the multivariate functional data, MFPCA is performed using the univariate functional principal components. The PCA basis functions are estimated from the $\mathbf{X}$. These basis functions are then applied to $n$ observations of $\mathbf{X}$ based on the PACE (PCA through conditional expectation) approach \citep{yao2005functional}. 

Let 
\[
\mathbf{X} = (X, Y, Z)^{\top},
\]
be a vector-valued stochastic process corresponding to the functional random variables related to the standardised landmark $x$, $y$ and $z$ coordinates, respectively.  
For $p \in \{x, y, z\}$, let $I_x$ be a compact set in $\mathbb{R}$ with finite Lebesgue measure such that 
\[
X : I_x \to \mathbb{R}, \quad X \in L^2(I_x),
\]
where $L^2(I_x)$ denotes the space of square-integrable functions on $I_x$. Similarly, $(I_y, L^2(I_y))$ and $(I_z, L^2(I_z))$ are defined.  

The $P$-fold Cartesian product is defined as
\[
I := I_x \times I_y \times I_z,
\]
so that $X$ is a stochastic process indexed by $t \in I$ and taking values in the product Hilbert space
\[
H := L^2(I_x) \times L^2(I_y) \times L^2(I_z).
\]

The inner product on $H$ is defined as:
\[
\langle\!\langle f, g \rangle\!\rangle := \sum_{p \in \{x, y, z\}} \langle f_p, g_p \rangle 
= \sum_{p \in \{x, y, z\}} \int_{I_p} f_p(t_p) g_p(t_p) \, dt_p,
\]
for $f = (f_x, f_y, f_z)^{\top}, g = (g_x, g_y, g_z)^{\top} \in H$. Then $(H, \langle\!\langle \cdot, \cdot \rangle\!\rangle)$ is a Hilbert space with induced norm $\| \cdot \|$ \citep{happ2018multivariate}.

\paragraph{Multivariate Karhunen--Lo\`eve Representation.}\mbox{}\\

Assume that 

\[
\mathbb{E}[\mathbf{X}(t)] := \big(\mathbb{E}[X(t_x)], \mathbb{E}[Y(t_y)], \mathbb{E}[Z(t_z)]\big)^{\top} = 0, 
\quad \forall t = (t_x, t_y, t_z)^{\top} \in I.
\]
Let $C$ denote the $3 \times 3$ matrix-valued covariance function:
\begin{equation}
C(s, t) = \mathbb{E}\big[ \mathbf{X}(s) \mathbf{X}(t)^{\top} \big],
\end{equation}
where the $(p,q)$th element of $C(s,t)$ is:
\[
C_{p,q}(s_p, t_q) = \mathbb{E}[X_p(s_p) X_q(t_q)] = \operatorname{Cov}(X_p(s_p), X_q(t_q)),
\]
with $s_p \in I_p, t_q \in I_q$; $p,q \in \{x, y, z\}$ and $X_x = X$, $X_y = Y$ and $X_z = Z$.  

Define $\Gamma : H \to H$ as the covariance operator of $\mathbf{X}$:
\[
(\Gamma f)^{(q)}(t_q) := \langle\!\langle C_{\cdot, q}(\cdot, t_q), f(\cdot) \rangle\!\rangle 
= \sum_{p=1}^P \int_{I_p} C_{p,q}(s_p, t_q) f_p(s_p) \, ds_p.
\]

By the spectral theorem for Hilbert--Schmidt operators, there exists an orthonormal basis $\{\varphi_j\}_{j=1}^{\infty} \subset H$ and eigenvalues $\lambda_1 \ge \lambda_2 \ge \dots \ge 0$ such that:
\[
\Gamma \varphi_j = \lambda_j \varphi_j, \quad \lambda_j \to 0 \ \text{as} \ j \to \infty.
\]
The multivariate Karhunen--Loève representation is then:
\[
\mathbf{X}(t) = \sum_{j=1}^{\infty} \xi_j \varphi_j(t), \quad t \in I,
\]
where $\xi_j = \langle\!\langle \mathbf{X}, \varphi_j \rangle\!\rangle$ are zero-mean uncorrelated random variables with $\operatorname{Cov}(\xi_j, \xi_l) = \lambda_j \delta_{jl}$.

The truncated expansion for $J \ge 1$ is:
\[
\mathbf{X}_{[J]}(t) = \sum_{j=1}^J \xi_j \varphi_j(t),
\]
and for the $p$-th component:
\[
X_{p, [J_p]}(t_p) = \sum_{j=1}^{J_p} \xi_{p,j} \varphi_{p,j}(t_p),
\]
where $\varphi_{p,j}$ are the univariate FPCA basis functions of $X_p$.

\paragraph{Multivariate Functional Principal Component Analysis in Practice} \mbox{}\\

In practice, the eigenvalues $\{\lambda_j\}$, eigenfunctions $\{\varphi_j\}$, and scores $\{\xi_j\}$ must be estimated from $n$ i.i.d. observations $\mathbf{X}^{(1)}, \dots, \mathbf{X}^{(n)}$ of $\mathbf{X}$:

\begin{enumerate}
\item For each $X_p$, estimate the covariance function:
\[
\hat{K}_p(s, t) = \frac{1}{n-1} \sum_{i=1}^n X_p^{(i)}(s) X_p^{(i)}(t),
\]
and obtain $\hat{\varphi}_{p,j}$ and $\hat{\xi}_{p,j}^{(i)}$ for $j = 1, \dots, J_p$.

\item Form the score matrix $\Xi \in \mathbb{R}^{n \times J}$ with $J = \sum_{p \in \{x, y, z\}} J_p$.

\item Estimate $Z \in \mathbb{R}^{J \times J}$ by:
\[
\hat{Z} = \frac{1}{n-1} \Xi^{\top} \Xi.
\]

\item Perform an eigen-decomposition of $\hat{Z}$ to get $\hat{\lambda}_j$ and $\hat{v}_j$.

\item Compute the estimated multivariate eigenfunctions:
\[
\hat{\varphi}_{p,j}(t_p) = \sum_{k=1}^{J_p} [\hat{v}_j]_{(p,k)} \, \hat{\varphi}_{p,k}(t_p).
\]
\end{enumerate}

\subsubsection{Square-Root Velocity Function}

In this approach, we apply the functional data morphometric (FDM) framework to \emph{Procrustes-aligned} landmark data, incorporating the square-root velocity function (SRVF) representation of landmark curves.  
This method treats each 3D landmark trajectory as a multivariate function, enabling \emph{elastic shape analysis} that respects the non-linear geometry of shape space.\\
The SRVF framework allows shape comparison by separating changes in \emph{geometry} (shape) from changes in \emph{parameterisation} (speed along the curve).  
In the FDM context, SRVF transforms 3D landmark trajectories into a form where elastic distances can be computed efficiently while maintaining translation invariance.  
Rotation invariance is ensured via GPA, and parameterisation invariance via arc-length reparameterisation.\\

Let each specimen be represented by a set of $N$ ordered landmarks in $\mathbb{R}^3$ over a parameter domain $T$.  
For specimen $i$, the landmark trajectory is:
\[
f^{(i)} : T \to \mathbb{R}^3,
\]
with
\[
f^{(i)}(t) = \big(x^{(i)}(t), \, y^{(i)}(t), \, z^{(i)}(t)\big)^{\top}.
\]

After GPA, translation and scale effects are removed, yielding aligned trajectories.
 
Following Srivastava et al.~\citep{srivastava2011registration}, the SRVF of a smooth curve $f$ is defined as:
\begin{equation}
q(t) = \frac{\dot{f}(t)}{\sqrt{\|\dot{f}(t)\|}}, \quad t \in [0,1].
\label{eq:srvf}
\end{equation}

The Square Root Velocity Function (SRVF) representation possesses key properties that make it valuable for shape analysis: it is naturally invariant to translations since differentiation removes any constant shifts; the $L^2$ distance between SRVFs corresponds to the elastic geodesic distance in shape space, enabling an efficient comparison of shapes through standard inner products. However, rotation and parameterisation invariances are not inherent and must be explicitly enforced. Rotation invariance is typically obtained by aligning SRVFs via Procrustes superimposition, while parameterisation invariance is achieved by reparameterizing all curves to a common domain using uniform arc-length parameterisation.

Once the SRVFs $\{q^{(i)}\}_{i=1}^n$ are obtained, they are treated as multivariate functional observations:
\[
Q^{(i)}(t) = \big(q_x^{(i)}(t), \, q_y^{(i)}(t), \, q_z^{(i)}(t)\big)^{\top}.
\]
These are then analysed using the FDM pipeline: conversion into a functional data object, smoothing, and application of multivariate functional PCA (MFPCA) for dimensionality reduction and shape variation analysis.





\section{Simulation Study} \label{sec:sim}


We conducted a simulation study to assess the innovations driven by functional data in the morphometric analysis of three-dimensional shapes. Specifically, we compared eight pipelines: \textit{GM}, \textit{arc-GM}, \textit{FDM}, \textit{arc-FDM}, \textit{elastic-SRV-FDM}, \textit{arc-elastic-SRV-FDM}, \textit{soft-SRV-FDM}, and \textit{arc-soft-SRV-FDM}.

\subsection{Simulation Design}
Four distinct groups were generated based on an initial helix configuration: Group~1 maintained the original helix, Group~2 applied sinusoidal perturbations to the \(x\)- and \(y\)-coordinates, Group~3 introduced a sinusoidal perturbation to the \(z\)-coordinate, while Group~4 incorporated a phase shift into the \(x\)- and \(y\)-coordinates, leaving the \(z\)-coordinate unaltered. The simulations involved four groups with sizes \((21,\,32,\,124,\,23)\), and their proportions were carefully selected to align with those observed in kangaroo data. Each curve was defined over \(t \in [0,1]\) and was sampled at \(n_{\mathrm{points}}=30\) uniformly spaced intervals,
\[
t_i = \frac{i-1}{n_{\mathrm{points}}-1}, \quad i=1,\dots,n_{\mathrm{points}}.
\]
The simulated coordinates are then:
\noindent\textbf{Group 1}
\begin{align*}
x(t) &= \sin(2\pi t), \\
y(t) &= \cos(2\pi t), \\
z(t) &= t.
\end{align*}

\noindent\textbf{Group 2:}
\begin{align*}
x(t) &= \sin(2\pi t) + 0.15\,\sin(6\pi t), \\
y(t) &= \cos(2\pi t) + 0.10\,\cos(4\pi t), \\
z(t) &= t.
\end{align*}

\noindent\textbf{Group 3:}
\begin{align*}
x(t) &= \sin(2\pi t), \\
y(t) &= \cos(2\pi t), \\
z(t) &= t + 0.20\,\sin(4\pi t).
\end{align*}

\noindent\textbf{Group 4:}
\begin{align*}
x(t) &= \sin(2\pi t + \phi), \\
y(t) &= \cos(2\pi t + \phi), \\
z(t) &= t,
\end{align*}
where \(\phi\) denotes a small phase shift applied to the helix.

\medskip
Each specimen underwent an independent monotone time-warp,
\[
t \mapsto t^{\,u}, \quad u \sim \mathrm{Unif}(0.8,1.2),
\]
to mimic realistic phase variation in curve parameterisation. This step reflects natural variability in biological shape data, where features were locally stretched or compressed, and tests the robustness of the pipelines to misalignment. In addition, independent and identically distributed Gaussian noise was added to each coordinate with group-specific standard deviations \((\sigma_{G1},\sigma_{G2},\sigma_{G3},\sigma_{G4})=(0.05,\,0.05,\,0.10,\,0.06)\). Simulations were performed over \(n_{\mathrm{reps}}=50\) independent replicates.

\paragraph{Pipeline Settings.}
\begin{itemize}
  \item Arc-length reparameterisation: 30 output points.
  \item Geometric morphometrics (GM): 30 equally-spaced pseudo landmarks aligned by GPA.
  \item FDM basis: cubic B-splines on \([0,1]\) with 10 basis functions.
  \item Multivariate FPCA (MFPCA): target \(M=30\) components with safe downgrading if needed.
  \item Component selection: smallest \(k\) such that cumulative variance explained \(\ge 0.95\).
  \item Elastic/SRVF alignment (for elastic and Soft-FDM variants): rotation- and scale-invariant, with \(\alpha_{\mathrm{soft}}=0.6\) and \(\lambda_{\mathrm{soft}}=0.01\).
\end{itemize}

\paragraph{Evaluation.}
We used a five-fold cross-validation scheme for evaluation. (Alignment, basis construction, and component extraction were recomputed separately for the training and test folds. Classifiers were fitted on scores from the training folds and applied to scores obtained from the corresponding test decompositions.) Classification was performed on the resulting principal component scores of each pipeline using:
\begin{itemize}
  \item linear discriminant analysis;
  \item multinomial logistic regression;
  \item linear SVM (untuned, with centering and scaling).
\end{itemize}

\subsection{Simulation results}

\begin{figure}[H]
  \centering
  \includegraphics[width=0.8\textwidth]{\detokenize{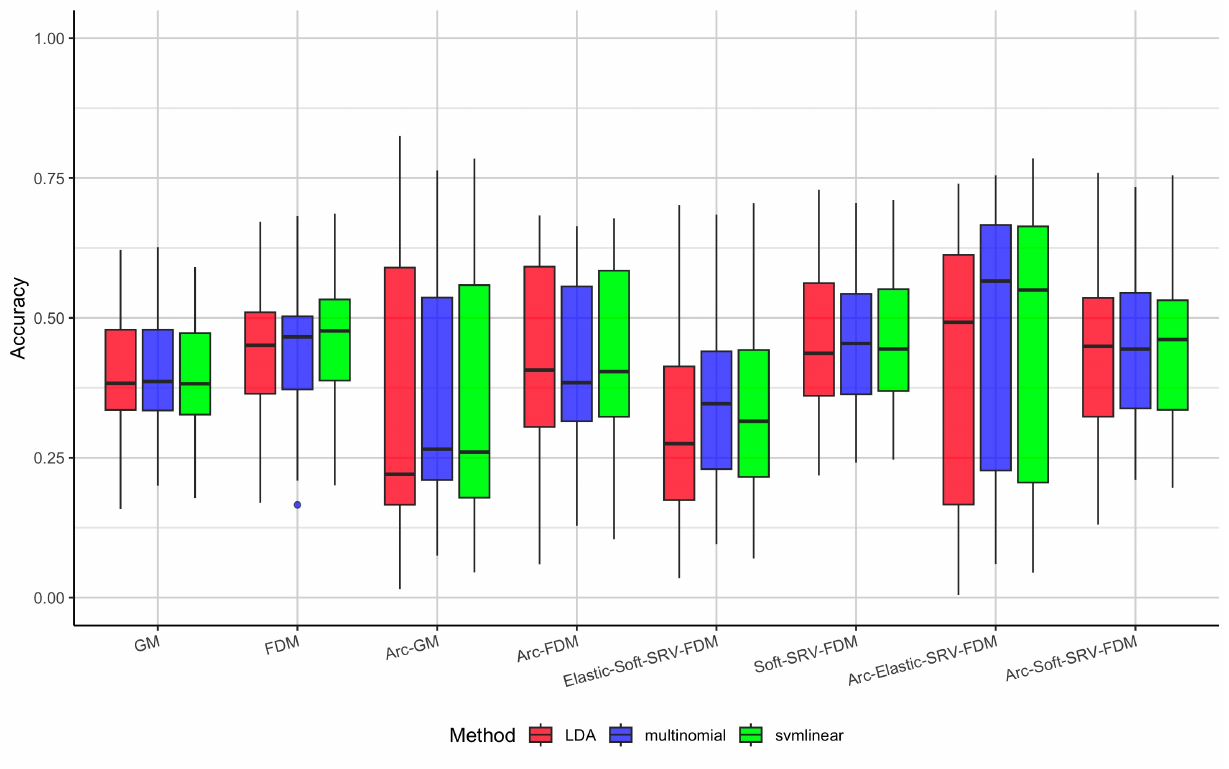}}
  \caption{Cross-validation accuracy across folds for the eight morphometric pipelines.} \label{fig:cv_accuracy}
\end{figure}
\begin{figure}[H]
  \centering
  \includegraphics[width=0.8\textwidth]{\detokenize{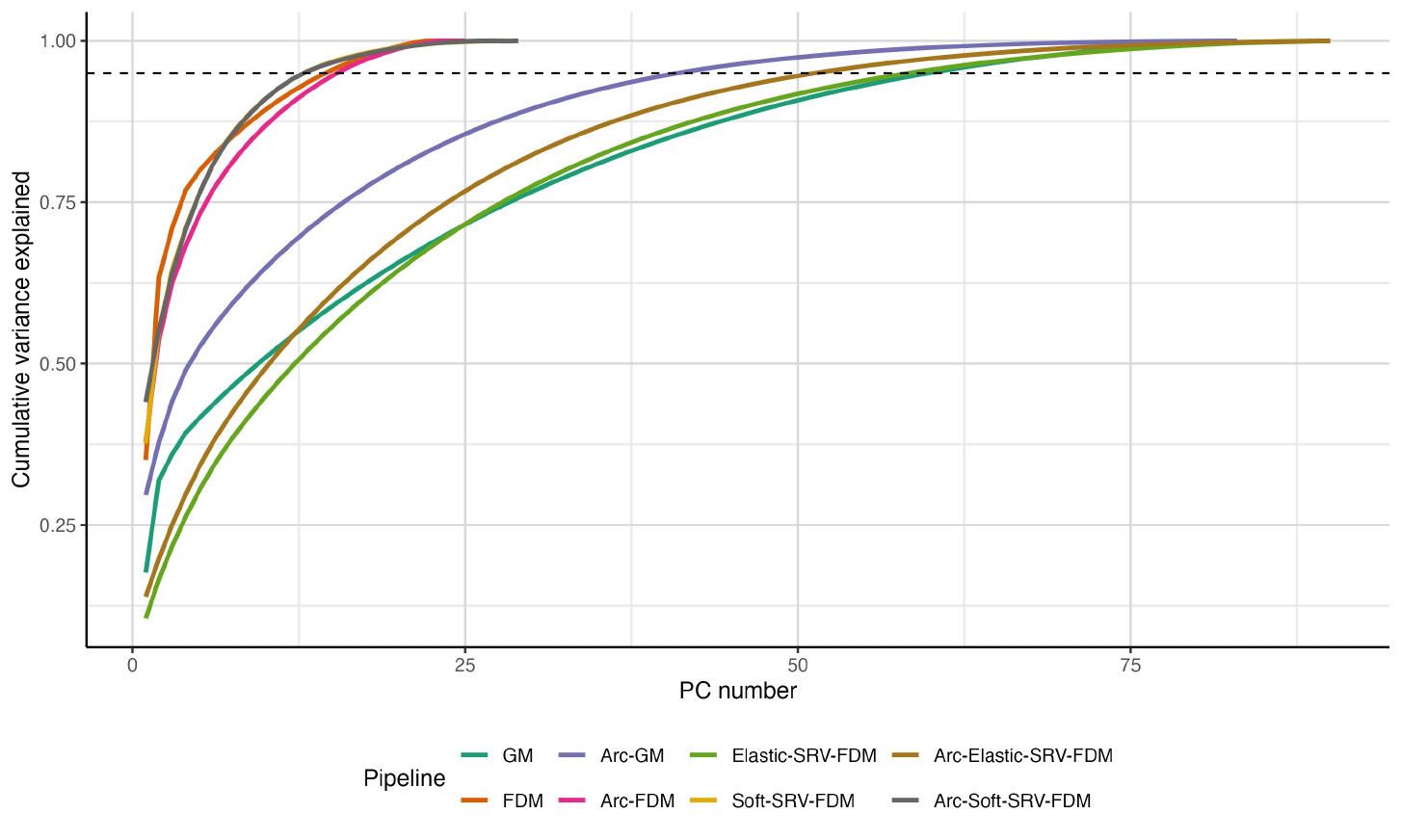}}
  \caption{Cumulative variance explained as a function of the number of components. The dashed line marks the 95\% cutoff.}\label{fig:cumvar}
\end{figure}
\begin{figure}[H]
  \centering
  \includegraphics[width=0.8\textwidth]{\detokenize{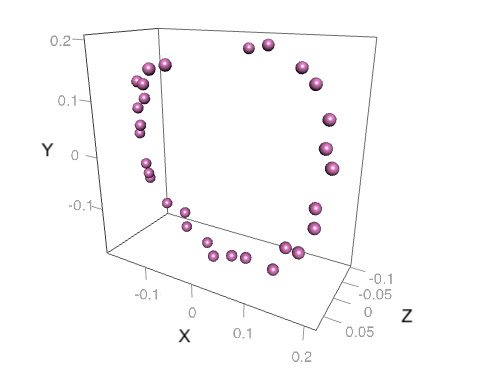}}
  \caption{Reconstruction landmark configuration of specimen 1 via arc-elastic-SRV-FDM (original, grey; reconstruction, pink).}\label{fig:recon_sim}
\end{figure}

Figure~\ref{fig:cv_accuracy} shows differences in classification performance, where functional data–driven approaches achieved higher predictive accuracy and better stability between folds. Figure~\ref{fig:cumvar}  illustrates that functional pipelines reached the 95\% variance threshold with fewer components than coordinate-based methods (Table~\ref{tab:pc95_simul}). The corresponding reconstruction plot (Figure~\ref{fig:recon_sim}) confirmed that the reconstructed shapes using the retained components preserved the curved structural information, thus validating the use of the 95\% cutoff. Both elastic and soft-aligned pipelines outperformed the purely coordinate-based and standard functional approaches. The elastic variants generally achieved good accuracy and efficient variance capture, but their performance showed greater variability across folds, indicating reduced stability. In contrast, soft-aligned pipelines offered consistently better accuracy with reduced variability and clearer group separation in PC space (Figure~\ref{fig::pc_plots}).

\begin{table}[H]
\centering
\caption{Number of principal components required to capture 95\% of cumulative variance across pipelines}\label{tab:pc95_simul}
\begin{tabular}{l c}
\hline
\textbf{Pipeline} & \textbf{Number of Principal Components} \\
\hline
GM & 60 \\
FDM & 14 \\
Arc-GM & 41 \\
Arc-FDM & 15 \\
Elastic-SRV-FDM & 58 \\
Soft-SRV-FDM & 13 \\
Arc-Elastic-SRV-FDM & 51 \\
Arc-Soft-SRV-FDM & 13 \\
\hline
\end{tabular}
\end{table}

\begin{table}[H]
\centering
\caption{Mean squared error of raw coordinates across pipelines (mean with standard deviation in parentheses).}
\begin{tabular}{l c}
\hline
Pipeline & MSE of Raw Coordinates \\
\hline
GM                  & \makecell{0.01994 \\ (0.00034)} \\
FDM                 & \makecell{0.00016 \\ (0.00081)} \\
Arc-GM              & \makecell{0.02089 \\ (0.00363)} \\
Arc-FDM             & \makecell{0.00425 \\ (0.00439)} \\
Elastic-SRV-FDM     & \makecell{0.00261 \\ (0.00112)} \\
Soft-SRV-FDM        & \makecell{0.00204 \\ (0.00081)} \\
Arc-Elastic-SRV-FDM & \makecell{0.00197 \\ (0.00126)} \\
Arc-Soft-SRV-FDM    & \makecell{0.00192 \\ (0.00120)} \\
\hline
\end{tabular}
\label{tab:mse_raw}
\end{table}

\begin{figure}[H]
  \centering
  \includegraphics[width=\textwidth,height=0.9\textheight,keepaspectratio]{\detokenize{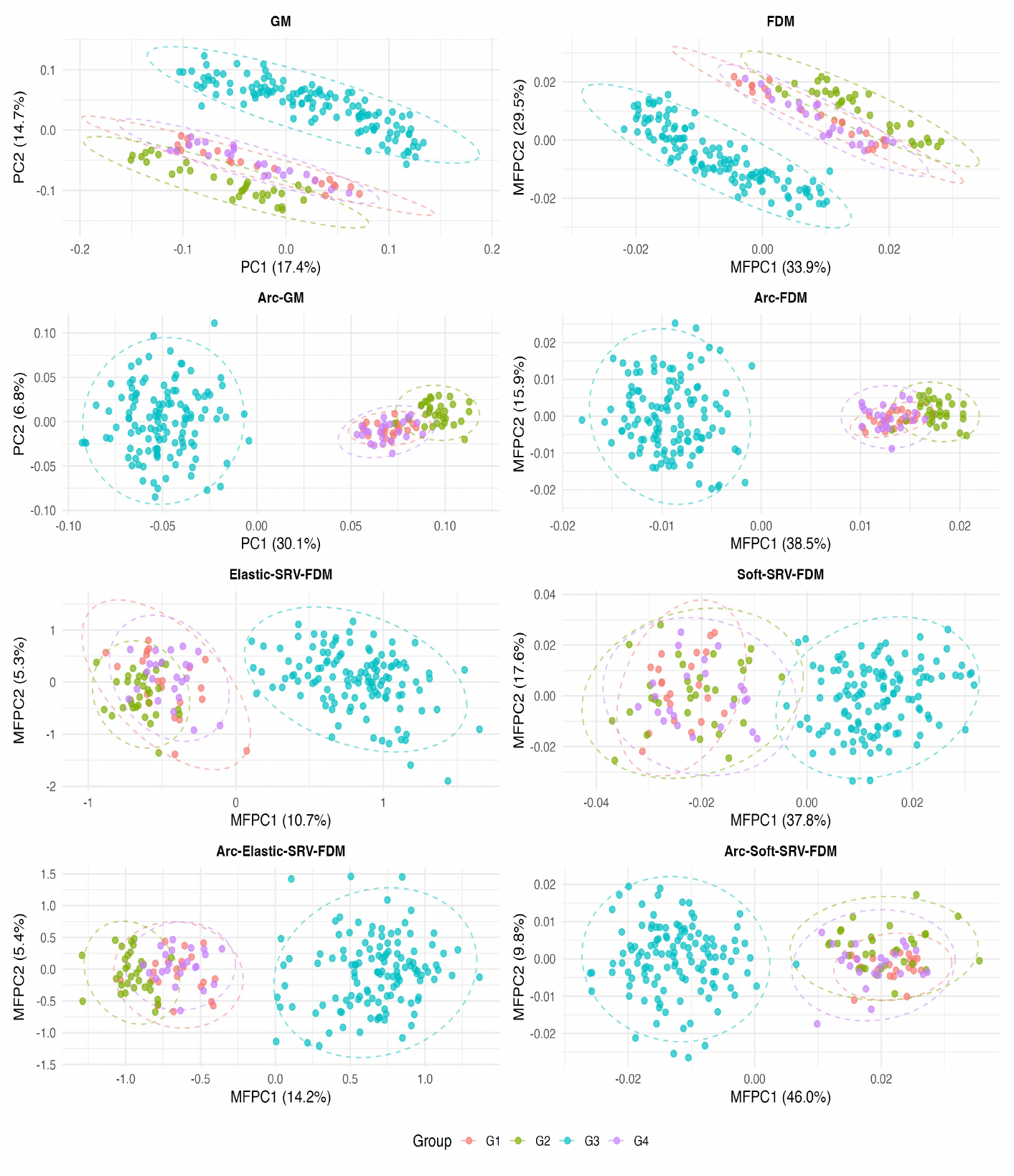}}
  \caption{PC1–PC2 projections illustrating group separation under the compared pipelines.}\label{fig:pc_plots}
\end{figure}

\section{Application to 3D Kangaroo Data} \label{sec:app}

\subsection{Kangaroo Cranial Landmark Dataset}

To analyse cranial shape variation in extant kangaroo species, we employed two parallel workflows: one based on GPA with sliding semi-landmarks, and another based on raw (unaligned) 3D landmark configurations. This dual approach allowed us to explore the effects of shape alignment and preprocessing on morphometric outcomes.

\subsubsection{Landmark Acquisition and Replicate Averaging}
A dataset of 3D cranial landmarks was obtained from the publicly available Butler Kangaroo Landmark collection \citep{butler20213d}. Each specimen was represented by 48 anatomical landmarks digitised across 41 extant specimens. Each specimen has been assigned with dietary category labels (browser, grazer, mixed feeder, omnivore) obtained from the classification file.

Each specimen was landmarked twice. Bland–Altman analysis showed negligible differences between replicates 1 and 2. Therefore, the aligned replicates were averaged to produce a single landmark configuration per specimen for further analysis.

The kangaroo dataset \citep{butler20213d} already includes a mixture of fixed landmarks and semi-landmarks digitised along cranial curves. These semi-landmarks were standardised using the established sliding procedure, which minimises bending energy relative to the Procrustes consensus. This ensured comparability with prior morphometric analyses of the dataset. 

To remove non-shape variation, raw landmark configurations were first subjected to GPA. For functional pipelines, landmark curves were interpolated and reparameterised either by time (landmark index) or by cumulative arc-length. The SRVF was then applied to enable elastic shape alignment, which was performed in both strict (elastic) and regularised (soft) forms. 

\begin{figure}[H]
  \centering
  \includegraphics[width=0.8\textwidth]{\detokenize{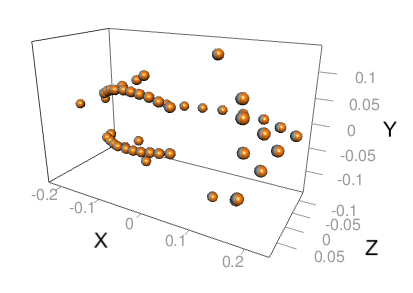}}
  \caption{Reconstruction landmark configuration of specimen 1 via elastic-SRV-FDM (original,
grey; reconstruction, orange).}\label{fig::reconplot}
\end{figure}

 Eight pipelines were implemented: GM, FDM, arc-GM, arc-FDM, elastic-SRV-FDM, soft-SRV-FDM, arc-elastic-SRV-FDM, and arc-soft-SRV-FDM. Each was subjected to PCA or MFPCA depending on the representation, and the number of retained principal components was determined by a 95\% cumulative variance threshold. The landmark configuration of a representative specimen (Specimen~1) was then reconstructed from the truncated components for each pipeline. In GM and Arc-GM, reconstruction added the rank-$k$ PCA approximation to the Procrustes consensus. In FDM and arc-FDM, the truncated MFPCA expansion from the functional representation was assembled across the $x$, $y$, and $z$ coordinates. For elastic and arc-Elastic pipelines, reconstruction was performed in SRVF space by adding the specimen’s truncated SRVF-PCA scores to the SRVF mean, followed by inverse SRVF mapping via time integration and resampling to the landmark grid. Soft-SRV-FDM and arc-soft-SRV-FDM used analogous truncated MFPCA predictions under soft alignment. Reconstructed shapes were Procrustes-superimposed to their corresponding originals for visual comparison, and reconstruction accuracy was quantified using mean squared error (Figure~\ref{fig::reconplot}). 
 
 To evaluate the discriminatory capacity of each pipeline, the retained PC scores were used as predictors in three classification methods: linear discriminant analysis (LDA), multinomial logistic regression, and linear-kernel support vector machines (SVM).The supervised learning methods are trained using the retained PC scores.

All analyses were conducted in R version 4.3.2)\cite{RStudioTeam2023}. Landmark-based morphometric procedures, including GPA, were implemented using the \texttt{geomorph}\cite{geomorph,geomorphadams2021}package. Functional data representations and elastic alignment were carried out using \texttt{fdasrvf}\cite{fdasrvf}, while MFPCA was performed with \texttt{MFPCA}\cite{mfpca} and \texttt{funData}\cite{fundata} packages. Classification models were fitted using \texttt{caret}\cite{caret}, which provided a unified framework for LDA, multinomial logistic regression, and SVM with linear kernels. Additional packages employed included \texttt{Morpho}\cite{morpho}, \texttt{MASS}\cite{mass}, \texttt{tidyr}\cite{tidyr}, \texttt{dplyr}\cite{dplyr}, \texttt{ggplot2}\cite{ggplot2}, and \texttt{rgl}\cite{rgl} for statistical computation, data wrangling, and visualisation.

\subsection{Results}

\begin{figure}[H]
  \centering
  \includegraphics[width=\textwidth]{\detokenize{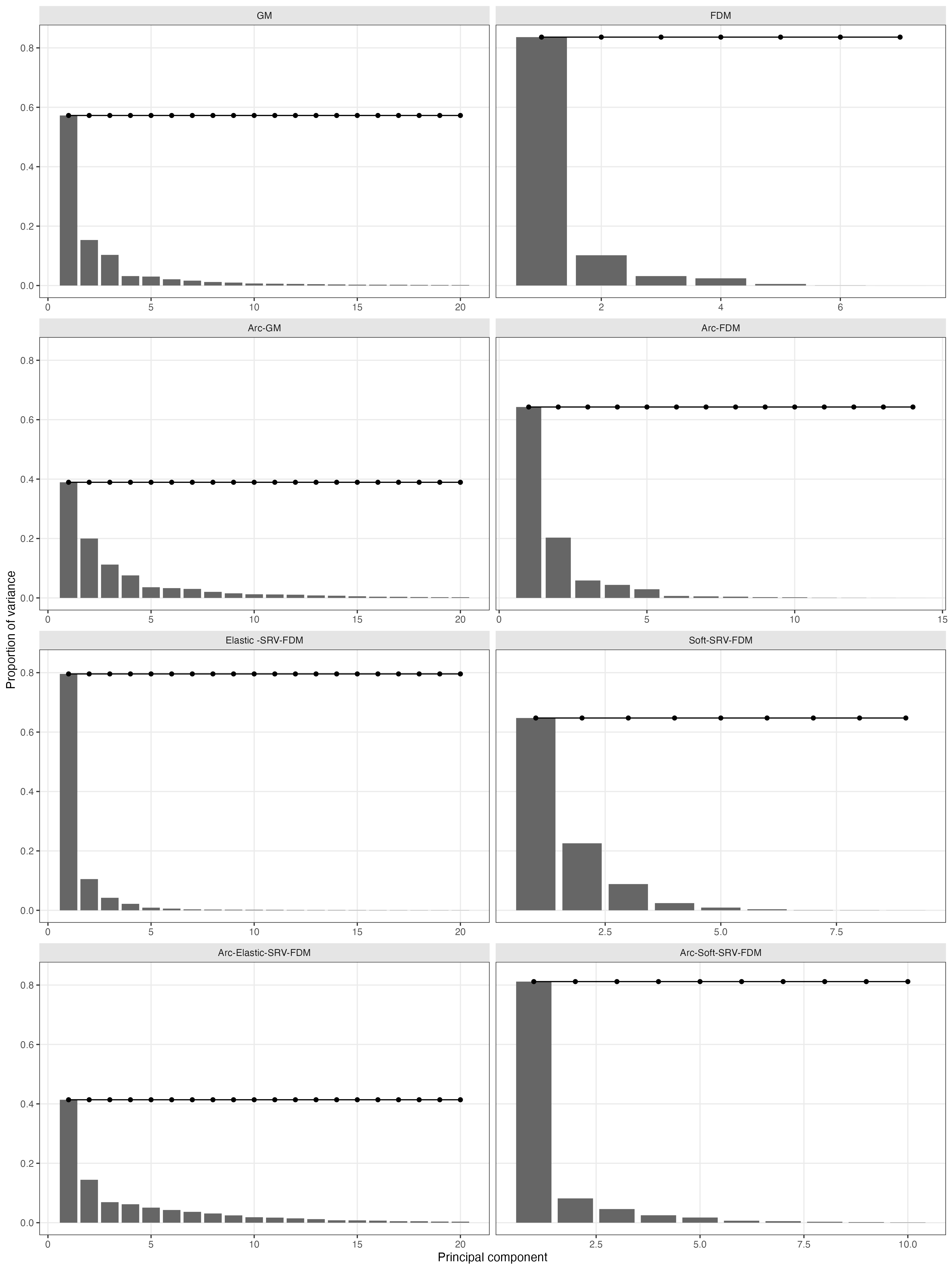}}
  \caption{Scree and cumulative variance plots for PCA and MFPCA pipelines.}\label{fig::screeplot}
\end{figure}

The number of principal components required to capture 95\% of the variance varied substantially across pipelines (Table~\ref{tab:pc95_kangaro}). In all cases, subsequent components contributed progressively smaller proportions of variance, with a sharp drop after the first two to three components (Figure~\ref{fig::screeplot}). This pattern was most pronounced in FDM and Soft-FDM, where the first three components alone accounted for nearly all of the variance. Similarly, arc-soft-FDM and elastic-SRV-FDM reached the 95\% threshold with only four components. By contrast, arc-GM and GM required more components, while arc-elastic-SRV-FDM  was the most component-intensive, requiring 15 components to achieve the same level of variance explanation.

\begin{figure}[H]
  \centering
  \includegraphics[width=\textwidth]{\detokenize{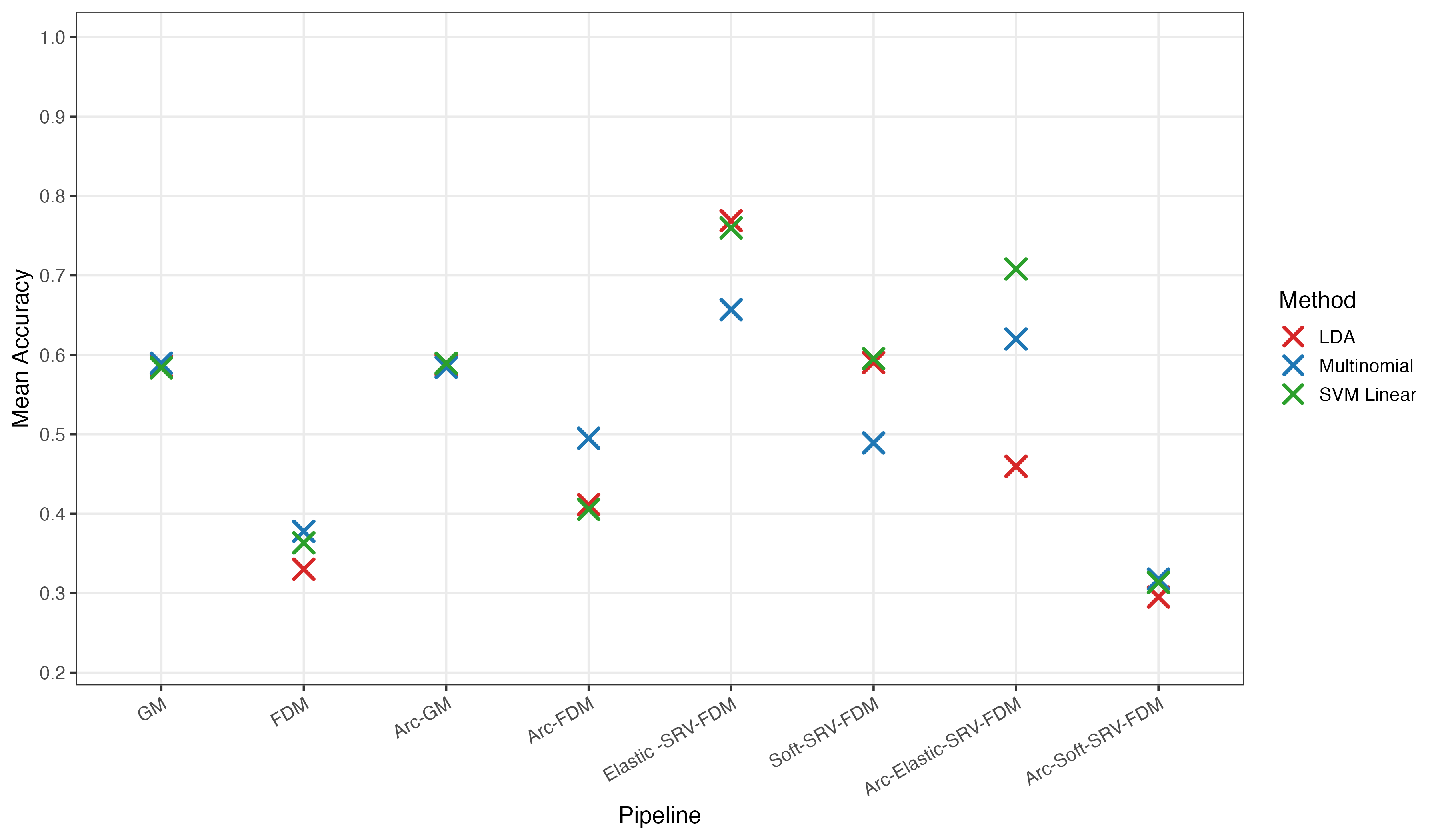}}
  \caption{Pipeline accuracy by classification methods.}\label{fig::pipe_accuracyplot}
\end{figure}

Classification accuracies varied across pipelines and methods. Elastic-SRV-FDM consistently outperformed other pipelines (Figure~\ref{fig::pipe_accuracyplot}), achieving the highest accuracy across classifiers, particularly with LDA. In contrast, GM, arc-GM and soft-SRV-FDM pipelines yielded comparable accuracies, while arc-soft-SRV-FDM yielded the lowest classification performance. 

\begin{figure}[H]
  \centering
  \includegraphics[width=\textwidth,height=3.0\textheight,keepaspectratio]{\detokenize{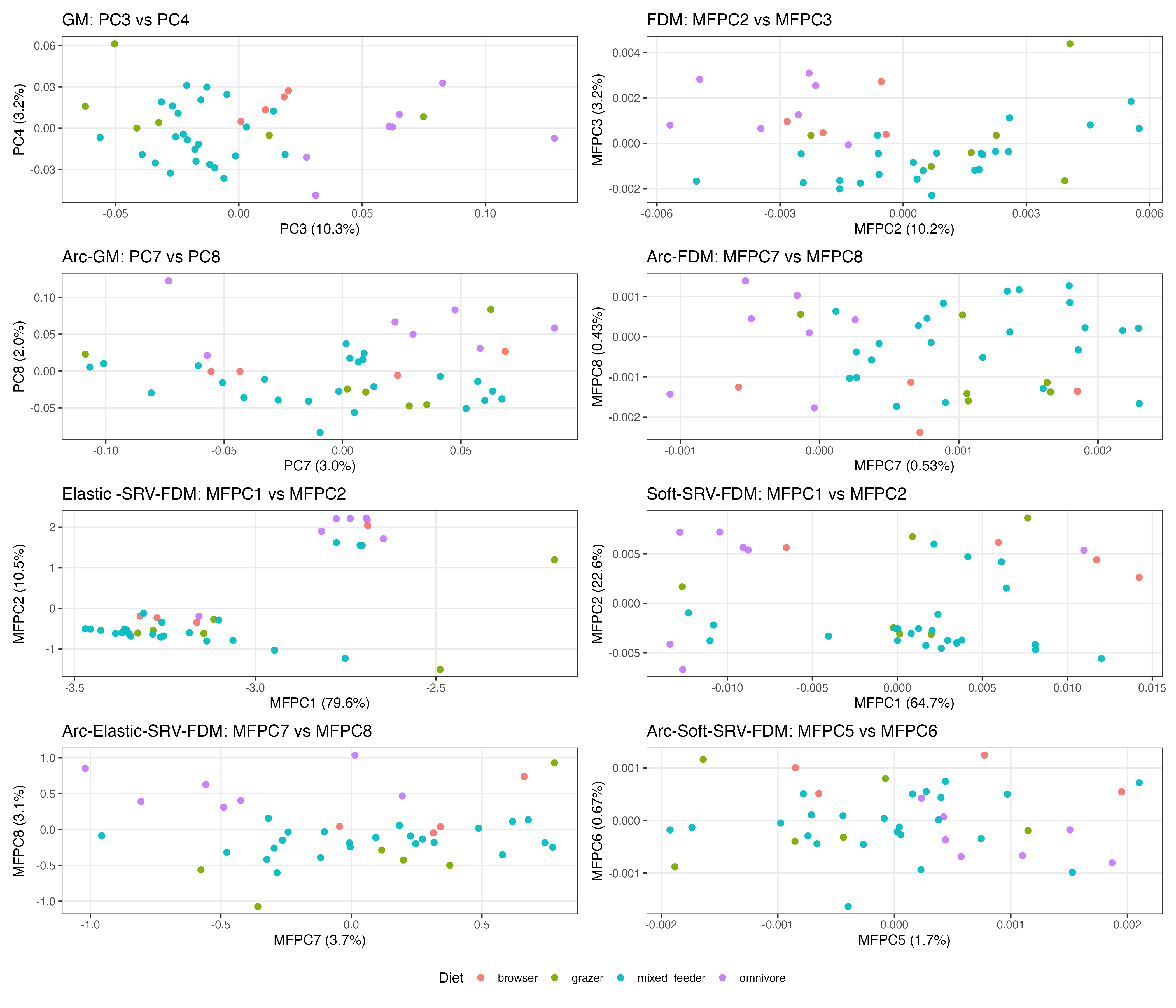}}
  \caption{Best adjacent PC/MFPC pairs by pipelines chosen via LDA cross-validation accuracy.}\label{fig::pc_plots}
\end{figure}

Plots of the best PC pairs for each pipeline revealed varying degrees of dietary group separation and within-group tightness (Figure~\ref{fig::pc_plots}. Overall, elastic-SRV-FDM and soft-SRV-FDM pipelines capture the largest proportion of shape variation, that is, over 85\% in the first two MFPCs. They also exhibit the most distinct groupings, especially for browsers and omnivores. While the clusters were slightly looser than those in the elastic-SRV-FDM analysis, browsers and omnivores again exhibited tighter cohesion, and the overlap between grazers and mixed feeders was reduced compared to GM and FDM.  In contrast, the arc-FDM and arc-soft-SRV-FDM pipelines relied on higher-order PCs that explained less than 2\% of the variance in each axis, revealed dispersed and intermixed. 

\begin{table}[H]
\centering
\caption{Number of principal components required to capture 95\% of cumulative variance across pipelines}\label{tab:pc95_kangaro}
\begin{tabular}{l c}
\hline
\textbf{Pipeline} & \textbf{Number of principal Components} \\
\hline
GM & 9 \\
FDM & 3 \\
Arc-GM & 13 \\
Arc-FDM & 5 \\
Elastic-SRV-FDM & 4 \\
Soft-SRV-FDM & 3 \\
Arc-Elastic-SRV-FDM & 15 \\
Arc-Soft-SRV-FDM & 4 \\
\hline
\end{tabular}
\end{table}

\begin{table}[H]
\centering
\caption{Mean squared error of kangaroo cranial coordinates across pipelines (mean with standard deviation in parentheses)}
\begin{tabular}{l c}
\hline
Pipeline & MSE of Raw Coordinates \\
\hline
GM                  & \makecell{0.01581 \\ (0.00001)} \\
FDM                 & \makecell{0.00299 \\ (0.00010)} \\
Arc-GM              & \makecell{0.01385 \\ (0.00068)} \\
Arc-FDM             & \makecell{0.01345 \\ (0.00069)} \\
Elastic-SRV-FDM     & \makecell{0.00424 \\ (0.00033)} \\
Soft-SRV-FDM        & \makecell{0.00481 \\ (0.00018)} \\
Arc-Elastic-SRV-FDM & \makecell{0.00841 \\ (0.00056)} \\
Arc-Soft-SRV-FDM    & \makecell{0.00794 \\ (0.00042)} \\
\hline
\end{tabular}
\label{tab:mse_raw2}
\end{table}

\section{Discussion}

Functional data morphometrics is advantageous because it captures smooth variation along entire curves or surfaces. The analysis of the kangaroo cranial dataset showed that the elastic SRV-FDM pipeline performed best, followed by the two GM pipelines. This outcome is expected, as the dataset is purely landmark-based, which inherently favours GM methods. Since curvature is not represented in such datasets, the advantages of functional data approaches are limited. An important property of the elastic SRVF framework is its ability to preserve amplitude variation while aligning phase \cite{srivastava2011registration}, which in this case is effectively synonymous with aligning the geometric structure of a landmark-only dataset. Consequently, the nature of this dataset does not favour soft-aligned functional data approaches, since the central idea of balancing the contributions of landmarks and continuous function shapes to reach an optimal registration \cite{guo2022data} cannot be realised when only landmarks are available.

The simulation study was designed to address these shortcomings by incorporating phase variation and moderate coordinate noise, to assess how the eight pipelines perform under noisy, imbalanced folds. The soft-alignment approach of Guo et al. (2022) \cite{guo2022data}  has an advantage in this context, serving as a compromise that pushes landmarks closer without requiring exact overlays, while balancing contributions of landmarks and functional shapes to achieve an optimal registration.

The combination of arc-length parameterisation with SRVF alignment also provides a powerful framework for morphometric pipelines. The SRVF framework (after arc-length normalisation) simplifies the originally complicated elastic metric into a straightforward $L^2$  product \cite{srivastava2011registration}. For the kangaroo data, arc-GM appears to perform better in terms of accuracy compared to its GPA-aligned GM counterpart, because arc-length parameterisation eliminates the influence of uneven landmark spacing or irregular sampling along the curve. When more balanced shapes are generated in the simulation study, the advantage of combining arc-length parameterisation and SRVF becomes evident, with pipelines implementing these aspects showing improved accuracy.

The soft-registration approach is likely to have an edge when images are processed from scratch, as it allows the relative contributions of functional shape and landmarks to be determined directly from the data itself \cite{guo2022data} . However, its effectiveness is not evident in the kangaroo dataset, but is better illustrated in the simulated data, since the former is purely landmark-based.

For reconstruction, the number of principal components required to capture 95\% of variation was used for each pipeline. The data application and simulation results in general show that pipelines combining arc-length parameterisation with SRVF alignment achieve strong reconstruction accuracy, as reflected by low mean MSE values (Table~\ref{tab:mse_raw} and Table~\ref{tab:mse_raw2}), thereby reinforcing the strength of elastic functional frameworks in handling phase variability. Notably, FDM also performs particularly well, yielding the lowest MSE and underscoring its efficiency in reconstruction tasks where smooth functional representation is advantageous. Principal component analysis and MFPCA maximise variance rather than discrimination. Consequently, the first two PCs capture the dominant variation, but it is important to note that this may not correspond to group separation but may reflect global size or noise. Discriminatory structure can reside in later PCs with smaller variance, as illustrated in the data application.

\section{Conclusion} \label{sec:conc} 

Our work compared landmark-based and functional data–driven pipelines for the analysis of three-dimensional cranial landmark data, using both simulations and real kangaroo cranial datasets to evaluate variance explained, reconstruction accuracy, and classification performance. Across both datasets, functional data approaches, and in particular arc-parameterised and SRVF-based methods, outperformed coordinate-based pipelines by preserving biologically meaningful structural information and improving dietary classification accuracy. While FDM yielded the strongest reconstruction accuracy, elastic SRVF-based methods provided superior classification accuracy, highlighting complementary strengths across pipelines. The present work was based on three-dimensional landmark coordinate data, which neglects curvature, where this limitation is addressed  through the simulated data. For future studies, it would be valuable to extend the analysis to full three-dimensional images that capture both raw coordinates and curvature, thereby enabling a more robust evaluation, particularly when applying soft-aligned and elastic pipelines. Arc-length parameterisation and SRVF alignment also show great potential for advancing morphometric frameworks, and these innovations could additionally be incorporated into our two-dimensional functional data morphometrics framework \cite{pillay2024functional} to enhance 2D analyses. Overall, these directions highlight the advantages of functional data analysis for complex morphological datasets, where elastic and soft alignment strategies, coupled with arc-length parameterisation, can better uncover ecologically meaningful patterns often obscured under traditional landmark-based approaches. The eight pipelines presented here provide a flexible framework for selecting methods best suited to the characteristics of the data.

\paragraph{Disclosure Statement} \mbox{}\\

During the preparation of this manuscript, the authors used ChatGPT (OpenAI) to improve the academic tone and enhance clarity of presentation. The tool was also used to reduce potential coding errors, improve readability, and ensure tidy code. All methodological design, analyses, and scientific interpretations were conducted solely by the authors. All content generated or suggested by the tool was critically reviewed, edited, and validated by the authors, who take full responsibility for the final version of the manuscript.

\bibliographystyle{abbrv}
\bibliography{ref}

\begin{thebibliography}{10}

\bibitem{geomorph}
D.~C. Adams, M.~L. Collyer, A.~Kaliontzopoulou, and E.~K. Baken.
\newblock Geomorph: Software for geometric morphometric analyses. r package version 4.0.10, 2025.

\bibitem{adams2013geomorph}
D.~C. Adams and E.~Ot{\'a}rola-Castillo.
\newblock geomorph: an r package for the collection and analysis of geometric morphometric shape data.
\newblock {\em Methods in ecology and evolution}, 4(4):393--399, 2013.

\bibitem{geomorphadams2021}
E.~K. Baken, M.~L. Collyer, A.~Kaliontzopoulou, and D.~C. Adams.
\newblock geomorph v4.0 and gmshiny: enhanced analytics and a new graphical interface for a comprehensive morphometric experience.
\newblock {\em Methods in Ecology and Evolution}, 12:2355--2363, 2021.

\bibitem{bhattacharya2012nonparametric}
A.~Bhattacharya and R.~Bhattacharya.
\newblock {\em Nonparametric inference on manifolds: with applications to shape spaces}, volume~2.
\newblock Cambridge University Press, 2012.

\bibitem{butler20213d}
K.~Butler, K.~J. Travouillon, A.~R. Evans, L.~Murphy, G.~J. Price, M.~Archer, S.~J. Hand, and V.~Weisbecker.
\newblock 3d morphometric analysis reveals similar ecomorphs for early kangaroos (macropodidae) and fanged kangaroos (balbaridae) from the riversleigh world heritage area, australia.
\newblock {\em Journal of Mammalian Evolution}, 28(2):199--219, 2021.

\bibitem{Drydmard98}
I.~L. Dryden and K.~V. Mardia.
\newblock {\em Statistical Shape Analysis}.
\newblock Wiley, Chichester, 1998.

\bibitem{Drydmard16}
I.~L. Dryden and K.~V. Mardia.
\newblock {\em Statistical Shape Analysis, with Applications in {R}. Second Edition.}
\newblock John Wiley and Sons, Chichester, 2016.

\bibitem{gunz2005semilandmarks}
P.~Gunz, P.~Mitteroecker, and F.~L. Bookstein.
\newblock {\em Semilandmarks in three dimensions}.
\newblock Springer, 2005.

\bibitem{guo2022data}
X.~Guo, W.~Wu, and A.~Srivastava.
\newblock Data-driven, soft alignment of functional data using shapes and landmarks.
\newblock {\em arXiv preprint arXiv:2203.14810}, 2022.

\bibitem{happ2018multivariate}
C.~Happ and S.~Greven.
\newblock Multivariate functional principal component analysis for data observed on different (dimensional) domains.
\newblock {\em Journal of the American Statistical Association}, 113(522):649--659, 2018.

\bibitem{fundata}
C.~Happ-Kurz.
\newblock Object-oriented software for functional data.
\newblock {\em Journal of Statistical Software}, 93(5):1--38, 2020.

\bibitem{mfpca}
C.~Happ-Kurz.
\newblock {\em MFPCA: Multivariate Functional Principal Component Analysis for Data Observed on Different Dimensional Domains}, 2022.
\newblock R package version 1.3-10.

\bibitem{hartlen2022arclength}
D.~C. Hartlen and D.~S. Cronin.
\newblock Arc-length re-parametrization and signal registration to determine a characteristic average and statistical response corridors of biomechanical data.
\newblock {\em Frontiers in Bioengineering and Biotechnology}, 10:843148, 2022.

\bibitem{huckemann2010intrinsic}
S.~Huckemann.
\newblock Intrinsic inference on the mean geodesic of planar shapes and tree discrimination by leaf growth.
\newblock {\em The Annals of Statistics}, 39(2):1024--1056, 2011.

\bibitem{katina2021functional}
S.~Katina, L.~Vittert, and A.~W.~Bowman.
\newblock Functional data analysis and visualisation of three-dimensional surface shape.
\newblock {\em Journal of the Royal Statistical Society Series C: Applied Statistics}, 70(3):691--713, 2021.

\bibitem{kent1997consistency}
J.~T. Kent and K.~V. Mardia.
\newblock Consistency of procrustes estimators.
\newblock {\em Journal of the Royal Statistical Society: Series B (Statistical Methodology)}, 59(1):281--290, 1997.

\bibitem{caret}
{Kuhn} and {Max}.
\newblock Building predictive models in r using the caret package.
\newblock {\em Journal of Statistical Software}, 28(5):1–26, 2008.

\bibitem{marron2021object}
J.~Marron and I.~Dryden.
\newblock {\em Object Oriented Data Analysis}.
\newblock Chapman and Hall/CRC, 2021.

\bibitem{rgl}
D.~Murdoch and D.~Adler.
\newblock {\em rgl: 3D Visualization Using OpenGL}, 2025.
\newblock R package version 1.3.24.

\bibitem{patrangenaru2015nonparametric}
V.~Patrangenaru and L.~Ellingson.
\newblock {\em Nonparametric Statistics on Manifolds and Their Applications to Object Data Analysis}.
\newblock CRC Press, Boca Raton, FL, 2015.

\bibitem{pillay2024functional}
A.~B. Pillay, D.~Pathmanathan, S.~Dabo-Niang, A.~Abu, and H.~Omar.
\newblock Functional data geometric morphometrics with machine learning for craniodental shape classification in shrews.
\newblock {\em Scientific Reports}, 14(1):15579, 2024.

\bibitem{RStudioTeam2023}
{R Core Team}.
\newblock {\em R: A Language and Environment for Statistical Computing}.
\newblock R Foundation for Statistical Computing, Vienna, Austria, 2023.

\bibitem{ramsay1998curve}
J.~O. Ramsay and X.~Li.
\newblock Curve registration.
\newblock {\em Journal of the Royal Statistical Society Series B: Statistical Methodology}, 60(2):351--363, 1998.

\bibitem{ramsay2005functional}
J.~O. Ramsay and B.~W. Silverman.
\newblock {\em Functional data analysis}.
\newblock Springer, 2005.

\bibitem{morpho}
S.~Schlager.
\newblock Morpho and rvcg -- shape analysis in {R}.
\newblock In G.~Zheng, S.~Li, and G.~Szekely, editors, {\em Statistical Shape and Deformation Analysis}, pages 217--256. Academic Press, 2017.

\bibitem{SrivastavaKlassen2016}
A.~Srivastava and E.~P. Klassen.
\newblock {\em Functional and Shape Data Analysis}.
\newblock Springer Series in Statistics. Springer New York, NY, 1 edition, 2016.

\bibitem{srivastava2011registration}
A.~Srivastava, W.~Wu, S.~Kurtek, E.~Klassen, and J.~S. Marron.
\newblock Registration of functional data using fisher-rao metric.
\newblock {\em arXiv preprint arXiv:1103.3817}, 2011.

\bibitem{fdasrvf}
J.~D. Tucker.
\newblock {\em fdasrvf: Elastic Functional Data Analysis}, 2025.
\newblock R package version 2.4.0.

\bibitem{Tumpach2017}
A.~B. Tumpach and S.~C. Preston.
\newblock Quotient elastic metrics on the manifold of arc-length parameterized plane curves.
\newblock {\em Journal of Geometric Mechanics}, 9(2):227--256, 2017.

\bibitem{mass}
W.~N. Venables and B.~D. Ripley.
\newblock {\em Modern Applied Statistics with S}.
\newblock Springer, New York, fourth edition, 2002.
\newblock ISBN 0-387-95457-0.

\bibitem{webster2010practical}
M.~Webster and H.~D. Sheets.
\newblock A practical introduction to landmark-based geometric morphometrics.
\newblock {\em The paleontological society papers}, 16:163--188, 2010.

\bibitem{ggplot2}
H.~Wickham.
\newblock {\em Ggplot2: Elegant graphics for data analysis}.
\newblock Use R! Springer International Publishing, Cham, Switzerland, 2 edition, June 2016.

\bibitem{dplyr}
H.~Wickham, R.~François, L.~Henry, and K.~Müller.
\newblock {\em dplyr: A Grammar of Data Manipulation}, 2023.
\newblock R package version 1.1.4.

\bibitem{tidyr}
H.~Wickham, D.~Vaughan, and M.~Girlich.
\newblock {\em tidyr: Tidy Messy Data}, 2024.
\newblock R package version 1.3.1.

\bibitem{wu2024shape}
Y.~Wu, C.~Huang, and A.~Srivastava.
\newblock Shape-based functional data analysis.
\newblock {\em Test}, 33(1):1--47, 2024.

\bibitem{yao2005functional}
F.~Yao, H.-G. M{\"u}ller, and J.-L. Wang.
\newblock Functional data analysis for sparse longitudinal data.
\newblock {\em Journal of the American statistical association}, 100(470):577--590, 2005.

\bibitem{zelditch2004geometric}
M.~L. Zelditch, D.~L. Swiderski, and H.~D. Sheets.
\newblock {\em Geometric Morphometrics for Biologists: A Primer}.
\newblock Elsevier Academic Press, San Diego, 2nd edition, 2004.

\end{thebibliography}

\end{document}